\documentclass[12pt,preprint]{emulateapj}

\usepackage{enumerate}

\slugcomment{}

\shorttitle{Photochemical enrichment of deuterium in Titan's atmosphere}

\shortauthors{Cordier et \textit{al.}}

\newcommand{\chfour}{$\mathrm{CH_{4}}${ }}
\newcommand{\chthreed}{$\mathrm{CH_{3}D}${ }}

\begin{document}


\title{Photochemical enrichment of deuterium in Titan's atmosphere: new insights from Cassini-Huygens}

\author{
Daniel~Cordier\altaffilmark{1,2},
Olivier~Mousis\altaffilmark{3,4},
Jonathan I.~Lunine\altaffilmark{3},
Audrey~Moudens\altaffilmark{1}
\& V{\'e}ronique~Vuitton\altaffilmark{3,5}
}

\altaffiltext{1}{Institut de Physique de Rennes, CNRS, UMR 6251, Universit{\'e} de Rennes 1, Campus de Beaulieu, 35042 Rennes, France}

\email{daniel.cordier@ensc-rennes.fr}

\altaffiltext{2}{Ecole Nationale Sup{\'e}rieure de Chimie de Rennes, Campus de Beaulieu, 35700 Rennes, France}

\altaffiltext{3}{Lunar and Planetary Laboratory, University of Arizona, Tucson, AZ, USA}

\altaffiltext{4}{Universit{\'e} de Franche-Comt{\'e}, Institut UTINAM, CNRS/INSU, UMR 6213, 
                 Observatoire des Sciences de l'Univers de Besan\c{c}on Cedex, France}

\altaffiltext{5}{Universit{\'e} Joseph Fourier, Laboratoire de Plan{\'e}tologie de Grenoble, CNRS/INSU, France}

\begin{abstract}
Cassini-Huygens data are used to re-examine the potential sources of the D/H enhancement over solar, measured in methane, in Titan's atmosphere. Assuming that the system is closed with respect to carbon, the use of constraints from the Huygens probe for the determination of the current mass of atmospheric methane and the most up-to-date determination of D/H from Cassini/CIRS infrared spectra allow us to show that photochemical enrichment of deuterium is not sufficient to be the sole mechanism yielding the measured D/H value. A possible fractionation between CH$_3$D and CH$_4$ during the escape process may slightly enhance the deuterium enrichment, but is not sufficient to explain the observed D/H value over the range of escape values proposed in the literature. Hence, alternative mechanisms such as a primordial  deuterium enrichment must be combined with the photochemical enrichment in Titan's atmosphere in order to explain its current D/H value. 

\end{abstract}


\section{Introduction}

The interpretation of the D/H ratio measured in methane in the atmosphere of Titan is not straightforward. All measurements indicate values substantially higher than the protosolar value, namely the value in the hydrogen of the solar nebula, but less than the values in water in the Earth's oceans (SMOW) and in comets (B{\'e}zard \textit{et al.} 2007). There is no general agreement on the cause of the enhancement. Indeed, methane on Titan could have originated in the molecular cloud or interstellar medium with a high D/H value. Deuterium in methane would have then isotopically exchanged with molecular hydrogen in the gas phase at the time of entrapment in solids condensed in Saturn's formation zone (Mousis \textit{et al.} 2002) to obtain the D/H value observed in Titan. 

On the other hand, Titan could have been accreted from solids condensed in an initially hot and dense Saturn's subnebula. In this picture, the methane incorporated in Titan would result from  CO and CO$_2$ gas phase conversions (Prinn \& Fegley 1981) and would present an almost solar D/H ratio at the time of its trapping in solids ultimately accreted by the satellite (Lunine et \textit{al.} 1999; Mousis \textit{et al.} 2002). Pinto \textit{et al.} (1986) and Lunine \textit{et al.} (1999) argued that, because the initial methane reservoir on Titan was likely larger than what is seen today in the atmosphere of Titan, the observed D/H enhancement could be the result of photochemical enrichment of deuterium through that isotope's preferential retention during methane photolysis (Pinto  \textit{et al.} 1986; Lunine \textit{et al.} 1999).  The D/H ratio acquired by the atmospheric methane of Titan would be progressively enriched with time via photolysis, until it reaches the value observed today.

Here, we reinvestigate the hypothesis of photochemical enrichment of deuterium in the atmosphere of Titan, in light of a number of recent Cassini-Huygens measurements. Pinto \textit{et al.} (1986) and Lunine \textit{et al.} (1999) estimated the current mass of methane in the atmosphere of the satellite by assuming that its molar fraction is uniform, whatever the altitude. In contrast, we use the methane mole fraction and density atmospheric profiles resulting from data collected by the Gas Chromatograph Mass Spectrometer (GCMS) and Huygens Atmospheric Structure Instrument (HASI) onboard the Huygens probe during its descent  in Titan's atmosphere to better constrain the actual mass of methane. 
Cassini data also have provided new constraints on the vertical mixing in the atmosphere of Titan, leading Yelle \textit{et \textit{al.}} (2008) to derive the existence of a CH$_{4}$ escape flux which is about one third the photolytic destruction rate of CH$_4$.  Here, we investigate the influence of this prodigious escape on the total fractionation between CH$_3$D and CH$_4$.

Moreover, we utilize the most recent determination of D/H obtained by B{\'e}zard \textit{et al.} (2007) from Cassini/CIRS infrared spectra, which indicates values of $1.32^{+0.15}_{-0.11} \times 10^{-4}$, substantially higher than those employed by Lunine et \textit{al.} (1999) ($7.75\pm 2.25 \times 10^{-5}$; Orton 1992). Since the determination of B{\'e}zard \textit{et al.} (2007) was obtained by fitting simultaneously the $\nu$$_6$ bands of both $^{13}$CH$_3$D and $^{12}$CH$_3$D and the $\nu$$_4$ band of CH$_4$ from precise information on the CH$_4$ mixing ratio and temperature profile in the stratosphere available from the Huygens descent and limb-viewing Composite Infrared Spectrometer (CIRS) measurements, we believe that this measurement is more reliable than the previous ones.

All these revisions, together with the use of updated rate coefficients for methane loss derived from a recent photochemical model (Vuitton \textit{et al.} 2008), allow us to show that the photochemical enrichment of deuterium is not efficient enough in the atmosphere of Titan to explain its current D/H value, even if the current atmospheric reservoir of methane is postulated to exist over 4.5 Gyr. We conclude that the D/H ratio in methane initially trapped in Titan was already higher than the protosolar value prior to its release in the atmosphere.


\section{Isotopic Enrichment Model}

Our photochemical enrichment model is derived from the formalisms of Pinto \textit{et al.} (1986) and Lunine \textit{et al.} (1999). We determine the deuterium enrichment that occurred following the ultimate outgassing event that gave existence to the current atmospheric methane of Titan. This event might have started 4.5 Gyr ago at the epoch of core overturn, giving birth to a reservoir of methane dense enough to allow its survival up until now. On the other hand, the current reservoir of methane might also have resulted from a much more recent outgassing event, as late as $\sim$ 0.6 Gyr ago based on evolution models (Tobie \textit{et al.} 2006) which are consistent with the surface cratering record (Lorenz \textit{et al.} 2007). We assume that the system (atmosphere and subsurface from which the methane outgasses) is open with respect to the dissociated hydrogen (either deuterium or protium) which is lost to space, and to carbon only when methane loss is considered in the upper atmosphere of Titan.

We define $R$ as the ratio of the total mass of methane expelled from the interior of Titan (and constituting the initial reservoir) to the current atmospheric mass of methane. Let be $N_{1}(t)$ and $N_{2}(t)$ the respective total column abundance of CH$_{4}$ and CH$_{3}$D on Titan at time $t$. With the condition $N_{1} \gg N_{2}$, the evolution of $N_{1}(t)$ and $N_{2}(t)$ is given by

\begin{eqnarray}
\label{eq_diff_1}
\frac{\mathrm{d}N_{1}}{\mathrm{d}t} & = & - \frac{N_{1}(t)}{N_{1}(t)+N_{2}(t)} F - \frac{N_{1}(t)}{N_{1}(t)+N_{2}(t)} \Phi \nonumber \\
& \simeq & -(F+\Phi)
\end{eqnarray}

\begin{eqnarray}
\label{eq_diff_2}
\frac{\mathrm{d}N_{2}}{\mathrm{d}t} & = & - q \frac{N_{2}(t)}{N_{1}(t)+N_{2}(t)} F - l \frac{N_{2}(t)}{N_{1}(t)+N_{2}(t)} \Phi \nonumber \\
& \simeq & -(q F + l \Phi) \frac{N_{2}(t)}{N_{1}(t)}
\end{eqnarray}

\noindent where $F$ is the net photolytic destruction rate of CH$_{4}$ and $\Phi$ the methane escape rate. The parameter $q$ corresponds to the ratio of $k_{2}$ over $k_{1}$, namely the respective rates for \chthreed and \chfour destruction, and ranges between 0.8 and 0.88 from consideration of the chemical kinetics of deuterated species (Lunine \textit{et al.} 1999). The value of the free parameter $l$ ranges between 0 and 1, representing a possible fractionation between CH$_{3}$D and CH$_{4}$ during the escape process.

The integration of Eq.~\ref{eq_diff_1} and Eq.~\ref{eq_diff_2} leads to the following expression for $R$:

\begin{equation}
\label{new_R}
\mathrm{Log} R = \frac{\displaystyle 1}{\displaystyle 1-\frac{\displaystyle q F + l \Phi}{\displaystyle F + \Phi}} \mathrm{Log} f,
\end{equation}

\noindent where $f$ is defined as the ratio of D/H observed in Titan's current atmospheric methane to protosolar D/H ((D/H)$_\odot$~=~$2.35 \pm 0.3 \times 10^{-5}$; Mousis \textit{et al.} 2002).

Alternatively, $R$ can be expressed as follows (Lunine \textit{et al.} 1999):

\begin{eqnarray}
\label{eq_R_1}
R & =  & \frac{\displaystyle x^{(s)}_{\rm CH_{4}} \, 
P^{(s)}/g + m_{\rm CH_{4}} \, (F + \Phi) \, \tau}{\displaystyle x^{(s)}_{\rm CH_{4}} \,P^{(s)}/g} \nonumber \\
& \simeq & \frac{\displaystyle g \, m_{\rm CH_{4}} \, (F + \Phi) \, 
\tau}{\displaystyle x^{(s)}_{\rm CH_{4}} \, P^{(s)}},
\end{eqnarray}

\noindent where $g$ is the Titan's surface gravity, $m_{\rm CH_{4}}$ the mass of a methane molecule, 
$x^{(s)}_{\rm CH_{4}}$ the methane surface mole fraction ($x^{(s)}_{\rm CH_{4}}$ = $4.9 \times 10^{-2}$; Niemann \textit{et al.} 2005) and $P^{(s)}$ the pressure at the surface of Titan. The time elapsed since the formation of the initial methane reservoir up to now is $\tau$. Since the term $x^{(s)}_{\rm CH_{4}} P^{(s)}/g$ corresponds to the total mass of methane per unit of area, $R$ can be written as:

\begin{equation}
\label{eq_R_2}
  R = \frac{\displaystyle m_{\rm CH_{4}} \, (F + \Phi) \, \tau}{\displaystyle M_{\rm CH_{4}}},
\end{equation}

\noindent  where $M_{\rm CH_{4}}$ is the cumulative mass of atmospheric methane per unit  area. 

We adopt here $F = 6.9 \times 10^{13}$ molecules $\mathrm{m^{-2}.s^{-1}}$, a value derived from the photochemical model of Vuitton  \textit{et al.} (2008). In addition, when the escape of CH$_4$ is considered in our calculations, we set $\Phi = 2.8 \times 10^{13}$ m$^{-2}$.s$^{-1}$, namely the value of CH$_4$ escape derived by Yelle \textit{et al.} (2008) from Cassini INMS measurements. Otherwise, $\Phi$ is set to 0 and the deuterium enrichment is calculated in a way similar to the work of Pinto \textit{et al.} (1986) and Lunine \textit{et al.} (1999).

\begin{figure}
\resizebox{\hsize}{!}{\includegraphics[angle=0]{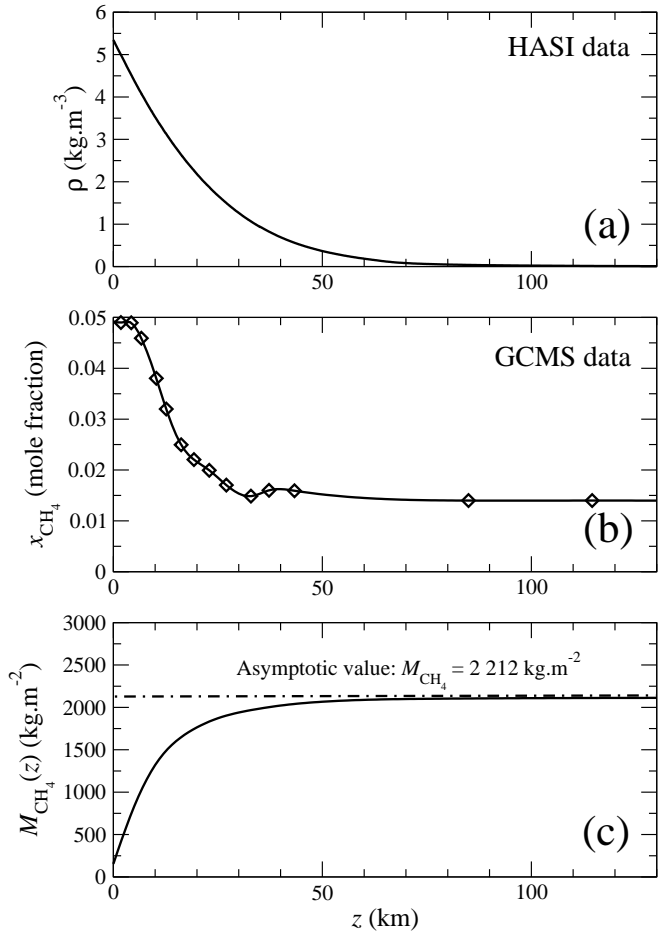}} 
\caption{(a) Density $\rho$ of Titan's atmosphere as a function of the altitude $z$ derived from the HASI data (Fulchignoni et \textit{al.} 2005). (b) Methane mole fraction $x_{\rm CH_4}$ as a function of the altitude derived from the GCMS data (Niemann et \textit{al.} 2005). Diamonds show the \textit{in situ} measurements and the solid line corresponds to the values interpolated with the spline technique. (c) Cumulative mass of methane in Titan's atmosphere as a function of the altitude.} 
\label{Mch4_xch4}
\end{figure}

Huygens probe data allow us to accurately determine the value of $M_{\rm CH_{4}}$. Indeed, knowing the atmospheric density and methane mole fraction profiles (see Figs. 1(a) and 1(b)), one can write:

\begin{equation}
\label{eq_Mch4}
  M_{\rm CH_{4}} = \int_{z=0}^{z=H} X_{\rm CH_{4}}(z) \, \rho(z) \, \mathrm{d}z,
\end{equation}

\noindent where $X_{\rm CH_{4}}$ is the methane mass fraction, $\rho(z)$ the density profile and $z$ the altitude. $H$ is the altitude above which $M_{\rm CH_{4}}$ has no significant change. Figure \ref{Mch4_xch4}(c) shows that $M_{\rm CH_{4}}$ $\simeq$ 2112 kg.m$^{-2}$ is asymptotically reached when the altitude exceeds $\sim$ 70 km. This determination is subtantially lower than the value of $\sim 5338$ kg.m$^{-2}$ derived from Lunine \textit{et al.} (1999). This discrepancy results from the use by Lunine \textit{et al.} (1999) (as well as by Pinto \textit{et al.} 1986) of a uniform methane mole fraction in the atmosphere of Titan whatever the altitude, in the absence of available $in~situ$ measurements.

\section{Results}
\label{results}

We consider two different cases of photochemical enrichment. In the first case, the CH$_4$ escape from the upper atmosphere of Titan is neglected.  In the second, this escape and a consequent additional fractionation between CH$_3$D and CH$_4$ are included.

\begin{figure}
\resizebox{\hsize}{!}{\includegraphics[angle=-90]{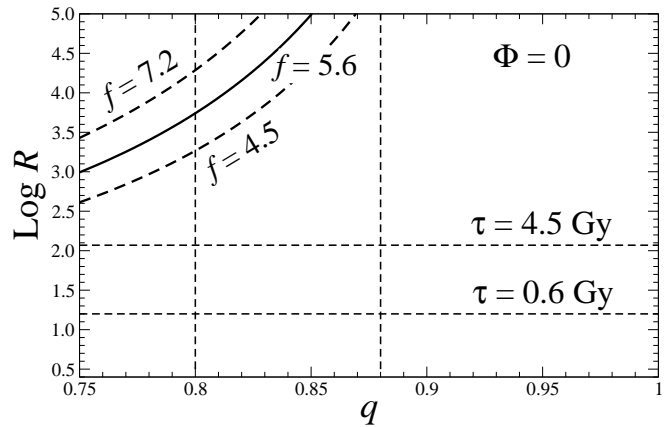}} 
\caption{The fractionation of deuterium in methane photochemistry with $\Phi$ = 0. Top left: plotted is $R$ as a function of $q$ (see Eq. \ref{new_R}). Three curves are shown, corresponding to different present-day deuterium enrichments measured in methane by Cassini/CIRS (see text). Bottom: the two horizontal lines represent values that would be acquired by $R$ for two different values of $\tau$ (see Eq. \ref{eq_R_2}). The two vertical lines represent limits on plausible values of $q$.}
\label{logR_q}
\end{figure}

Figure~\ref{logR_q} summarizes the results for deuterium enrichment via photodissociation calculated with Eq. \ref{new_R} assuming that $\Phi$ = 0, and shows the initial methane reservoir $R$ (normalized to the present one) against $q$. Three cases of present-day deuterium enrichment in the atmospheric methane of Titan are represented: the solid curve corresponds to the nominal value reported by B{\'e}zard \textit{et al.} (2007) ($f$ = 5.6) and the dashed curves to the extreme values ($f$ = 4.5--7.2) obtained when uncertainties are taken into account. We assume in all our calculations that the D/H ratio in the methane initially acquired by Titan during its accretion is protosolar. 

Note that a smaller value of $q$ yields greater fractionation for a given amount of methane photolysis, because it corresponds to deuterium being more tightly bound (Lunine \textit{et al.} 1999). Therefore, for a given deuterium enrichment, smaller values of $q$ allow the initial methane reservoir of Titan to be smaller than in the case required for higher values of $q$. The two horizontal lines represent values that would be acquired by $R$ if the actual methane reservoir were to exist since 0.6 or 4.5 Gyr ago, respectively (see Eq. \ref{eq_R_2}). Figure \ref{logR_q} shows that the initial reservoir of Titan's atmospheric methane was $\sim$ 16 times more massive than the current one if it was formed 0.6 Gyr ago, provided that $F$ remained fixed to its current value throughout the existence of this reservoir. If methane were present in the atmosphere of Titan since 4.5 Gyr ago, the initial reservoir was $\sim$ 126 times more massive than the current one. 

\begin{table}[h]
\caption[]{Deuterium enrichment expected in the methane of Titan's current atmosphere.}
\begin{center}
\begin{tabular}{lccc}
\hline
\hline
\noalign{\smallskip}
$\Phi$(m$^{-2}$.s$^{-1}$) & $\tau$				& $q$ = 0.8		& $q$ = 0.88\\
\hline
\noalign{\smallskip}
0 								& 0.6 Gyr  				& 1.7 			& 1.4 \\
0								& 4.5 Gyr  				& 2.6 			& 1.8 \\
2.8 $\times 10^{13}$ 				& 0.6 Gyr  				& 1.6 			& 1.3 \\
2.8 $\times 10^{13}$ 				& 4.5 Gyr  				& 2.1 			& 1.5 \\
\hline
\end{tabular}
\end{center}
\label{res}
\end{table}

Table \ref{res} summarizes the deuterium enrichments via photolysis in the atmospheric methane calculated for the adopted limits on plausible values of $q$ and the two different values of $\tau$, in the case $\Phi$ = 0. This table shows that, assuming a protosolar D/H in the methane originally released into the surface-atmosphere system, the photochemical enrichment is not efficient enough to allow the atmospheric D/H to reach the observed enrichment, even if the reservoir is postulated to have existed since 4.5 Gyr. A higher D/H ratio than the protosolar value must be advocated in the methane of Titan prior its outgassing, in order to explain the observed enrichment. Depending on the adopted value for $q$, the range of initial deuterium enrichments $f_{0}$ needed by the initial methane reservoir to allow photolysis to reach the nominal value of B{\'e}zard \textit{et al.} (2007) is between 3.2 and 4.0 for $\tau$ = 0.6 Gyr, and between 2.2 and 3.2 for $\tau$ = 4.5 Gyr. 
 
\begin{figure}
\resizebox{\hsize}{!}{\includegraphics[angle=-90]{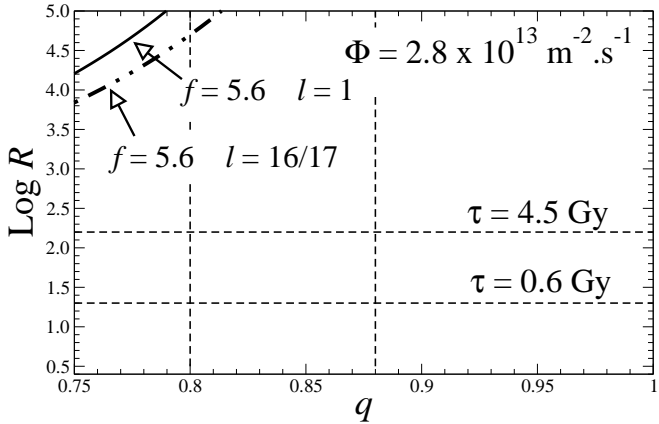}} 
\caption{Same as Fig. \ref{logR_q}, but with $\Phi = 2.8 \times 10^{13}$ m$^{-2}$.s$^{-1}$. Top left: both curves show $R$ as a function of $q$ in the case of $f$ = 5.6. The solid curve corresponds to the case where no fractionation is assumed during the methane escape ($l$ = 1). The dot-dot-dashed curve corresponds to the case where a fractionation between CH$_3$D and CH$_4$ occurs.}
\label{logR_q2}
\end{figure}

Figure~\ref{logR_q2} shows the results for deuterium enrichment via photodissociation calculated with Eq. \ref{new_R} assuming that $\Phi = 2.8 \times 10^{13}$ m$^{-2}$.s$^{-1}$, and shows $R$ against $q$. The two upper left curves have been determined for $f$ = 5.6 in Eq. \ref{new_R}, namely the nominal value reported by B{\'e}zard \textit{et al.} (2007). The solid curve corresponds to the case where no fractionation occurs between CH$_3$D and CH$_4$ during methane escape, while the dot-dot-dashed curve assumes there is fractionation. A word of caution must be given here. We have arbitrarily set the escape rate equal to the CH$_4$/CH$_3$D molar ratio ($l$ = 16/17) in order to quantify the influence of a possible fractionation between these two species. This is an order-of-magnitude estimate because the escape mechanism of methane remains unclear (Yelle \textit{et al.} 2008). As a result, the fractionation may behave differently. Figure ~\ref{logR_q2} shows that, despite the higher values of $R$ obtained with the methane loss in the upper atmosphere of Titan ($R$ $\simeq$ 22 for $\tau$ = 0.6 Gyr and $R$ $\simeq$ 166 for $\tau$ = 4.5 Gyr), the discrepancy between the values of $R(q)$ and $R(\tau)$ increases when compared with Fig. \ref{logR_q}, even if a fractionation between CH$_3$D and CH$_4$ is taken into account. This translates into a lower photochemical enrichment than when methane escape is neglected (see Table \ref{res}). The range of $f_{0}$ needed by the initial methane reservoir to allow the matching of photolysis with the nominal value of B{\'e}zard \textit{et al.} (2007) is consequently higher than without methane escape. Here, $f_0$ is between 3.6 and 4.3 for $\tau$ =  0.6 Gyr, and between 2.7 and 3.6 for $\tau$ =  4.5 Gyr, depending on the adopted value of $q$ and with no fractionation. If the fractionation between CH$_3$D and CH$_4$ is considered, the range of $f_0$ values is slightly lower, but still higher than the one inferred in the case of no methane escape.

\section{Discussion} 

Our results differ from those obtained by Lunine \textit{et al.} (1999) who concluded that the deuterium enrichment via photolysis was efficient enough to explain the current D/H value observed in Titan's atmosphere even if the methane incorporated in the forming satellite had acquired only a slightly supersolar D/H. By considering the Cassini-Huygens data, we show that the minimum value required for $f_0$ is higher than in Lunine \textit{et al.} (1999), even when the escape of methane and a possible fractionation between CH$_3$D and CH$_4$ are included. Therefore, additional mechanisms must be considered in order to explain the D/H value observed in Titan's methane. 

One possibility is isotopic thermal exchange of molecular hydrogen with \chthreed in the gas phase of the solar nebula. In this case, \chthreed might have originated from highly deuterium-enriched ices that vaporized when entering the nebula (Mousis \textit{et al.} 2002). During the cooling of the nebula, methane was trapped in crystalline ices around 10 AU, perhaps clathrates formed at $\sim$ 60 K, and incorporated into planetesimals that were preserved in the Saturnian subnebula because the latter was too cold to fully vaporize the ices (Mousis \textit{et al.} 2002; Alibert \& Mousis 2007). Subsequent to outgassing, additional enrichment in D/H would have occurred by the photochemical and escape mechanisms quantified here. 

Atreya \textit{et al.} (2006) have argued that methane formed in Titan's interior by reaction of CO and CO$_2$ with water and rock, in which case the D/H ratio in methane outgassed to the atmosphere would have been determined by that in the water, modified by fractionation associated with these so-called ``serpentinization'' reactions. Our results show that such a model for methane's origin must be able to produce a substantially supersolar value of D/H in the outgassed methane. Computation of the D/H fractionation associated with serpentinization depends on the details of the temperature in Titan's interior, the original D/H in the water, and possibly on the composition of the silicate buffer. It is unlikely to be large in view of the requisite high temperatures well above the water melting point) in Titan's interior. Thus, for this case, the considerations we made above for primordial methane would have to be invoked instead for the water, namely, that it was significantly enriched in D/H in the planetesimals that formed Saturn's largest moon. Finally, because the original appearance of methane in Titan's atmosphere could have been quite early, some 4.5 billion years ago, its origin might be tightly connected to the formation of the N$_{2}$ atmosphere from the dissociation of primordial NH$_{3}$. Hence, important amounts of H and H$_{2}$ could be present in the ancient atmosphere, leading to the possibility of additional fractionation between 
CH$_{4}$ and CH$_{3}$D via the following reversible  equations CH$_{4}$ $+$ HD $\rightleftharpoons$ CH$_{3}$D + H$_{2}$ and C $+$ CH$_{3}$ $=$ H + CH$_{2}$D (see Lee \textit{et al}. 2001 for a list of reactions forming back CH$_{4}$ and CH$_{3}$D from CH$_{3}$). However, it is difficult to quantify the influence of these reactions  on the final CH$_{3}$D/CH$_{4}$ ratio because the existence of a H$_{2}$-rich  primordial atmosphere remains uncertain.

\vspace{-1cm}

\begin{acknowledgements}
This work was supported in part by the CNES and NASA's Cassini program. We thank F. Ferri and M. Fulchignoni for providing us with the atmospheric density profile of Titan as measured by the Huygens probe.
\end{acknowledgements}


\end{document}